\documentclass[a4paper, conference]{IEEEtran}
\IEEEoverridecommandlockouts
\usepackage{cite}
\usepackage{tabularx}
\usepackage{amsmath,amssymb,amsfonts}
\usepackage{algorithmic}
\usepackage[bookmarks=false,hidelinks]{hyperref}
\usepackage{graphicx}
\usepackage{textcomp}
\usepackage{xcolor}
\usepackage{hyperref}
\usepackage{fancyhdr}
\usepackage{kantlipsum}
\fancypagestyle{plain}{
  \fancyhf{}
  \fancyhead[C]{Conference on \LaTeX}     
  \fancyfoot[L]{This is a notice}

}
\usepackage{eso-pic}

\usepackage[top=0.75in, bottom=1.05in, left=0.8in, right=0.8in]{geometry}
\def\BibTeX{{\rm B\kern-.05em{\sc i\kern-.025em b}\kern-.08em
    T\kern-.1667em\lower.7ex\hbox{E}\kern-.125emX}}

\begin{document}
\AddToShipoutPictureBG*{%
  \AtPageUpperLeft{%
    \setlength\unitlength{1in}%
    \hspace*{\dimexpr0.5\paperwidth\relax}
    \makebox(0,-0.75)[c]{\textit{Accepted for publication at the 7\textsuperscript{th} IEEE International Conference on Internet of Things and Intelligence System (IOTAIS 2024)}}
}}

\title{Threshold-Based Automated Pest Detection System for Sustainable Agriculture\\
\thanks{$^*$Joint first authors; Corresponding author: Tianle Li.}
\thanks{
This work was supported by the Microsoft FarmVibes project, and was developed as part of the TECHIN-515: Hardware Software Laboratory 2 course project at the Global Innovation Exchange, University of Washington, Seattle, WA, USA.}
}

\author{
    \IEEEauthorblockN{Tianle Li\textsuperscript{*}, Jia Shu\textsuperscript{*}, Qinghong Chen\textsuperscript{*}}
    \IEEEauthorblockA{
        \textit{Global Innovation Exchange} \\
        \textit{University of Washington}\\
        Seattle, WA, USA \\
        \{litianle, jshu3796, qic8\}@uw.edu}
    \and
    \IEEEauthorblockN{Murad Mehrab Abrar}
    \IEEEauthorblockA{
        \textit{Department of Mechanical Engineering} \\
        \textit{University of Washington}\\
        Seattle, WA, USA \\
        mabrar@uw.edu}
    \and
    \IEEEauthorblockN{John Raiti}
    \IEEEauthorblockA{
        \textit{Global Innovation Exchange} \\
        \textit{University of Washington}\\
        Seattle, WA, USA \\
        jraiti@uw.edu}
}

\maketitle

\begin{abstract}
This paper presents a threshold-based automated pea weevil detection system, developed as part of the Microsoft FarmVibes project. Based on Internet-of-Things (IoT) and computer vision, the system is designed to monitor and manage pea weevil populations in agricultural settings, with the goal of enhancing crop production and promoting sustainable farming practices. Unlike the machine learning-based approaches, our detection approach relies on binary grayscale thresholding and contour detection techniques determined by the pea weevil sizes. We detail the design of the product, the system architecture, the integration of hardware and software components, and the overall technology strategy. Our test results demonstrate significant effectiveness in weevil management and offer promising scalability for deployment in resource-constrained environments. In addition, the software has been open-sourced for the global research community.

\end{abstract}


\section{Introduction}
Agriculture is fundamental to human survival, yet pest management remains a significant challenge, particularly for small-scale farmers in resource-constrained environments \cite{barbedo2020detecting}. The primary objective of this paper is to develop an automated system for the timely detection and reporting of pea weevils, as even a single suspected pea weevil can be destructive to pea yields \cite{dore1995farm}. Traditional pest management techniques are often labor-intensive and require constant monitoring, which is impractical for many farmers. Our automated pest detection system leverages IoT and computer vision technologies to provide affordable and efficient weevil monitoring solutions, and an opensource software stack is available at \cite{githubGitHubKrantLeeee515MicrosoftFarmBeats}.

The system is designed to be cost-effective and open-source, making it suitable for impoverished, remote, and small farm areas. It offers remote monitoring capabilities, reducing the need for farmers to physically inspect their fields frequently. By accurately detecting weevil quantities, our system helps farmers to control pesticide usage and increase crop yields. Unlike the conventional pest control methods, our approach does not aim to kill pests but to detect them, allowing them to return to the field through a window at the bottom of the device. This design reduces the need for frequent device maintenance, thereby saving farmers valuable time. Also, the system enables farmers to monitor pest levels and make informed decisions about pest control measures. By addressing the challenges faced by farmers and agricultural professionals, the paper aims to enhance food security and promote sustainable farming practices.

\section{Related Work}

IoT-based systems have seen extensive application in agriculture for real-time monitoring and data collection. However, most implementations have focused primarily on environmental sensors to monitor variables such as humidity, temperature, and soil moisture, which are indirect indicators of pest activities \cite{st2018evaluation, farooq2021iot}. Pheromone attraction traps are often used in farms to attract and kill pests \cite{reddy2018pheromone, reddy2010new}, but the primary focus of our work is to attract and monitor weevil population to make decisions regarding pesticide usage. 

The literature contains various data-driven and machine learning-based approaches for pest management \cite{meisner2016data, durgabai2018pest}. Additionally, Convolutional Neural Networks (CNNs) are used in livestock monitoring to enhance animal health and farm management \cite{jung2021deep, huang2022deep}. However, these approaches often rely on large datasets, which are both challenging to acquire and labor-intensive to label, limiting their practical applicability in resource-constrained environments. Computer vision is widely applied in agriculture, such as in assessing the quality of fruits and vegetables, which automates the inspection process and provides more consistent assessments compared to manual inspection \cite{patricio2018computer, tian2020computer}. 

Despite these successes, detecting smaller pests like pea weevils remains challenging due to their small size and the presence of similar-looking debris in the environment. We address these challenges by using a precision camera with grayscale thresholding and contour detection techniques. Such thresholding techniques have been introduced in detecting sensor anomalies in autonomous vehicle fields \cite{abrar2023anomaly}, but a new contribution in pest detection. Also, Digital Signal Processing (DSP) have been applied in various domains for detecting defects, cracks, and  irregularities in materials by analyzing acoustic signals. Similarly, DSP can be used in agriculture to process images, identify pests based on specific signal patterns, enhance contrast, and remove noise--- thereby motivating us to develop a threshold-based pest detection system. 

\begin{figure*}[htbp]
\vspace{0.5cm}
\centerline{\includegraphics[width= 15cm]{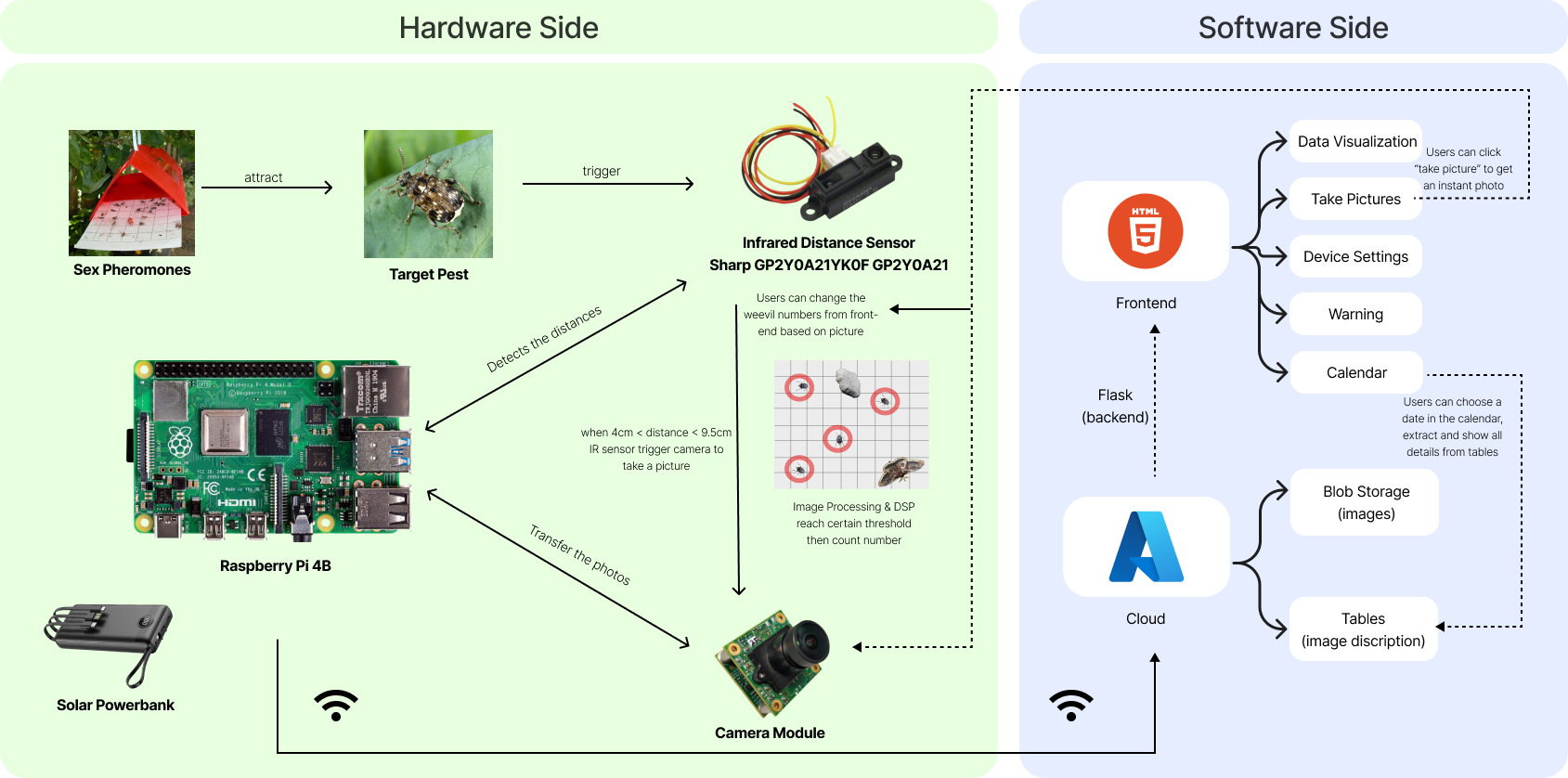}}
\caption{Hardware and software architecture of the threshold-based weevil detection system}
\label{fig-hwsw}
\end{figure*}

\section{Methodology}

\subsection{Theory}

The theory behind the developed threshold-based weevil detection system relies on three main components: (i) Binary Grayscale Thresholding, (ii) Image Comparison, and (iii) Contour Detection. 

\subsubsection{Binary Grayscale Thresholding}
The Binary Grayscale Thresholding technique is applied to enhance the contrast between weevils and the background. A function $f$ is defined as the weevil grayscale value function, which takes the pixel values of the image as input: $f(I)$ = $I(x, y)$; where $x, y$ are the pixel values.

Threshold $T$ is a grayscale value that is determined by empirical analysis. The binary thresholding function $D$ is then defined as:

\begin{equation*}
D(I) = 
\begin{cases}
    \textbf{0} \hspace{1cm} if \quad f(I) > T\\
    \textbf{255} \hspace{1.3cm} \textit{otherwise}
\end{cases}
\end{equation*}

where $D(I)$ converts pixels with grayscale values above $T$ to black (0) and those below $T$ to white (255).

\subsubsection{Image Comparison}
If a previous image exists, the system compares it with the current image using the absolute difference. The absolute difference $\Delta{I}$ is calculated as: 

\begin{equation*}
\Delta{I} = \big | I_t - I_{t-1} \big |    
\end{equation*}

Where $I_t$ and $I_{t-1}$ represent the current and previous grayscale images, respectively. A similarity threshold $S$ is computed based on empirical analysis. The system proceeds with further analysis based on the distance between $\Delta{I}$ and similarity threshold $S$.

\subsubsection{Contour Detection} 

The system uses computer vision techniques to identify and count contours in the image that fall within a specific area range, indicative of weevil size. Let $A$ represent the area of the detected contour. The contour detection function $C$ is defined as:

\begin{equation*}
C(I) = 
\begin{cases}
    \textbf{Weevil} & if \quad \textit{Lower Value} \leq A \leq \textit{Upper Value} \\
    \textbf{Not Weevil} & \hspace{2.4cm} \textit{otherwise}
\end{cases}
\end{equation*}

Based on this theory, the weevil detection system is developed. Our goal is to empirically find and tune the predefined grayscale threshold $T$, the similarity threshold $S$, and the lower and the upper pixel values for contour detection. 

\begin{figure*}[htbp]
\vspace{0.5cm}
\centerline{\includegraphics[width= 17cm]{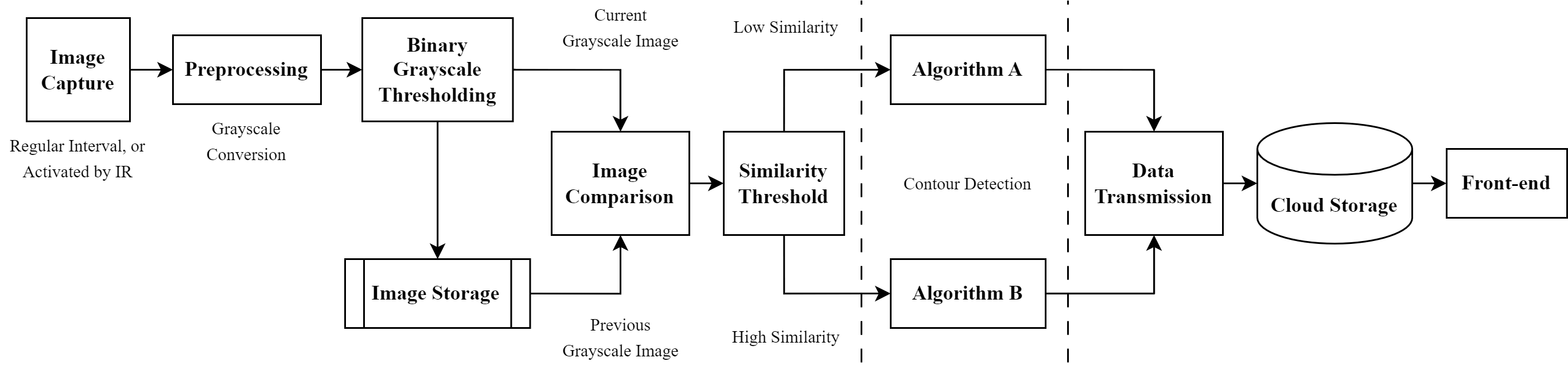}}
\caption{Function flow of the threshold-based weevil detection system}
\label{fig-flowchart}
\end{figure*}

\subsection{System Architecture}

The system comprises both hardware and software components designed to work together. The overall architecture is divided into two main sections: the hardware side and the software side, as illustrated in Fig. \ref{fig-hwsw}.

\subsubsection{Hardware}

The hardware of the weevil detection system comprises three key components: (i) a Raspberry Pi 4B Microprocessor, (ii) a Sharp Infrared (IR) Distance Sensor, and (iii) a Camera Module.

The Raspberry Pi 4B microprocessor functions as the central processing unit of the weevil detection system. It receives trigger signals from the Sharp IR sensor, which monitors the detection area. Sex pheromones are utilized to attract weevils into this area. Once a weevil enters, the IR sensor detects its presence and signals the Raspberry Pi to activate the camera module. The camera captures high-resolution images, with its auto-focus feature ensuring that each image is clear and detailed. The Raspberry Pi then processes the captured images and communicates with the backend for further analysis. A solar-powered battery supplies the necessary power to the Raspberry Pi and its connected components, ensuring continuous operation.





\subsubsection{Software}

The front end of the weevil detection system was developed using HTML and deployed on Microsoft Azure. The Azure backend handles data storage and processing, which includes blob storage for images and tables for metadata and image descriptions. DSP and image processing are conducted within the Raspberry Pi. When the Sharp IR sensor triggers the camera, the captured image undergoes DSP and image processing to check if pests meet a detection threshold. If they do, the image is transferred to the Azure backend for storage and further analysis. Users can monitor the system remotely through the front end, which includes features like homepage, image capture, settings, warnings, and calendar.

\subsection{Function Flow}

Fig. \ref{fig-flowchart} illustrates the function flow of the threshold-based weevil detection system. The following steps detail the overall process: 

\begin{itemize}
    \item Image Capture: The system uses a Raspberry Pi-controlled camera to capture images of the detection area (trap interior) at regular intervals or when triggered by the IR sensor.
    
    \item Preprocessing: Captured images are converted to grayscale to simplify the analysis and focus on luminance variations that indicate weevils.

    \item Binary Grayscale Thresholding: The grayscale thresholding technique is applied where pixels with a grayscale value below $T$ are converted to white (255), and those above $T$ are converted to black (0). This enhances the contrast between weevils and the background. We determine the value of $T$ from experimental analysis. 

    \item Image Comparison: If a previous image exists, the system compares it with the current image using absolute difference $\Delta{I}$. A similarity threshold $S$ is pre-defined. Comparing with this similarity threshold, the program decides which algorithm to use (A or B) to calculate the contours in the next step. 

    \item Contour Detection: The system uses computer vision to identify and count the contours in the image that fall within a specific area range, indicative of weevil size. Contours with areas between 
    \textit{Lower Value} and 
    \textit{Upper Value} pixels are considered weevils.

    \item Data Transmission: The processed data, including weevil count and image metadata, is uploaded to Azure for storage and further analysis, while alerts go to the farmer's app (Front-end). 
    
\end{itemize}

The above steps ensure that the system effectively differentiates weevils from other objects based on size and grayscale thresholding, thereby providing accurate and reliable pest detection to enhance agricultural productivity and sustainability.

\subsection{Algorithms}

The developed weevil detection system comprises two parts: (i) a primary algorithm, referred to as \textit{Algorithm A}, which counts all qualifying objects in the camera's field of view, and (ii) a supplemental algorithm, referred to as \textit{Algorithm B}, which counts only new qualifying objects in the camera's field of view.

\subsubsection{Primary Algorithm A}

Algorithm A performs counter-detection on all objects within the image after applying binary grayscale thresholding. While the application of machine learning models is common in image recognition tasks, we chose to employ a combination of DSP and computer vision techniques to identify weevils. This decision was driven by several critical factors, such as:

\begin{itemize}
    \item Lack of Suitable Public Datasets: Currently, there are no publicly available datasets that meet our specific requirements for training a machine learning model. Existing datasets typically consist of high-resolution images captured with microfocus lenses, where weevils occupy a significant portion of the image area. However, images automatically captured in the field with fixed cameras lack such angles and clarity, rendering these datasets unsuitable for our needs.

    \item Challenges in Dataset Creation: Creating a reliable dataset would necessitate capturing numerous weevil images in the field and meticulously labeling them, a process that is both time-consuming and costly. Additionally, weevil species can vary significantly across different regions, further complicating the dataset creation process.

    \item Cost-effectiveness and Practicality: After evaluating the accuracy and cost-effectiveness of using a machine learning model, we determined that a simpler, lower-cost, and more user-friendly approach would be more appropriate. Combining DSP and computer vision provides a practical solution that meets our project goals without the complexities associated with machine learning.
    
\end{itemize}

%

\subsubsection{Supplemental Algorithm B}

This algorithm focuses solely on detecting new objects that appear in the image after binary grayscale thresholding. In practice, we found that insects often die within our device, despite pathways designed to induce them to leave. In such cases, the dead insects remain within the camera's view. If we only use Algorithm A, these dead insects would be repeatedly counted as live insects, causing excessive false alerts and reducing device efficiency. This would force users to frequently clean the device, contradicting our design goal of saving farmers' time and effort.

Therefore, we employ absolute difference from computer vision as the basis for Algorithm B. After each photo capture, the system compares the current image with the previous one. A high similarity score indicates no significant changes in the device, allowing the algorithm to focus on newly detected objects. The system generally uses Algorithm A when the similarity is below threshold $S$ and shifts to Algorithm B when it exceeds this threshold.




\section{Experiment}


During the development of the threshold-based weevil detection system, we conducted two experiments to determine the optimal threshold values that provide accurate results. The system utilizes an IR sensor to trigger a camera for photographing weevils, conserving power by activating the camera only when weevils are detected. To assess the IR sensor’s efficiency, we also conducted another experiment to examine the impact of pea weevil body size and flying speed on the sensor’s accuracy. These experiments evaluate the performance of the system under various real-world conditions.


\subsection{Experiment 1: Determining the Grayscale Threshold T}

\begin{figure}[htbp]
  \centering
  \includegraphics[width=0.49\textwidth]{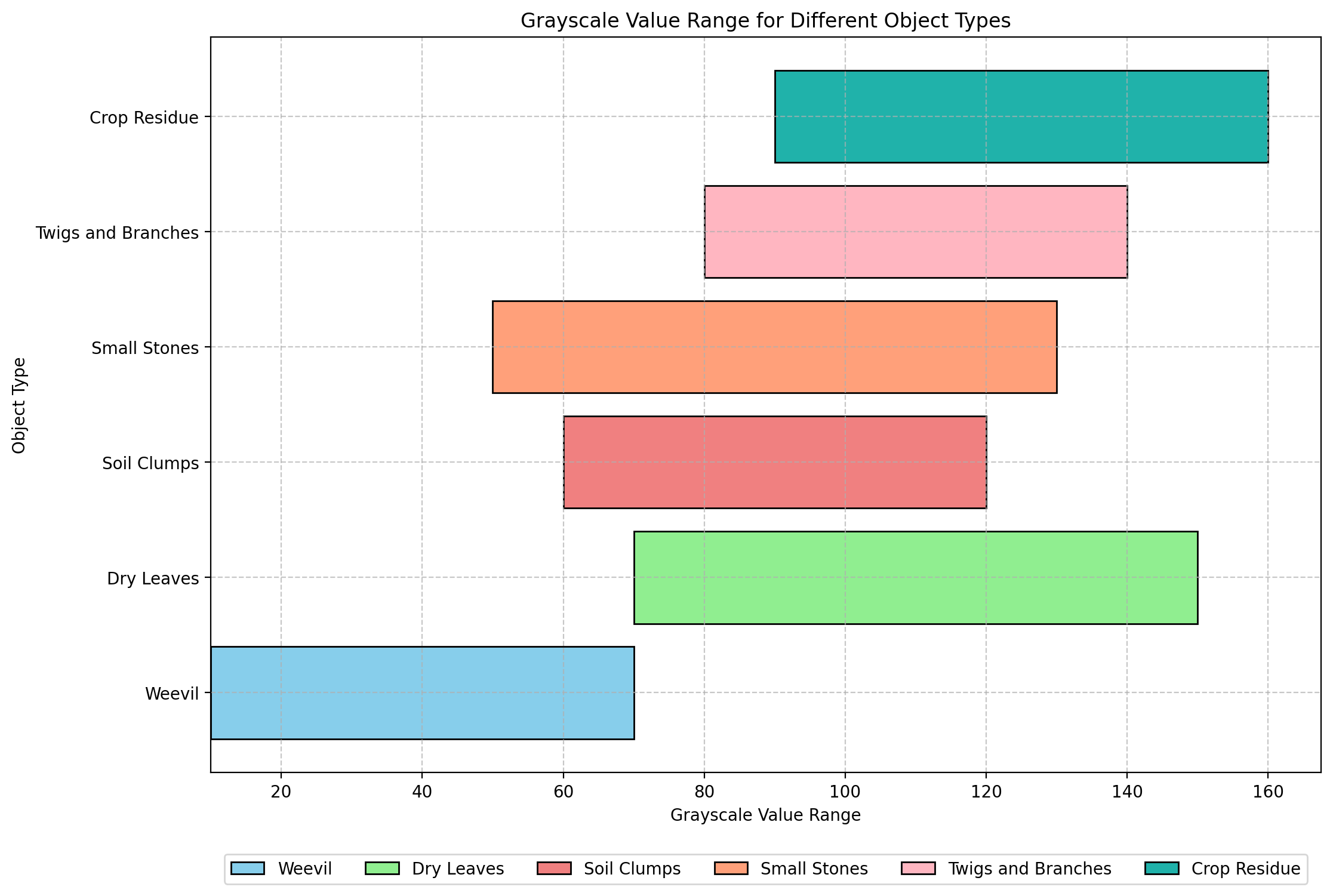}
  \vspace{2pt}
  \caption{Determining the Grayscale Threshold}
  \label{grayscale threshold}
\end{figure}

We conducted a statistical analysis of 100 images of different weevil species and potential interfering objects in the field to examine the grayscale value ranges and determine a suitable grayscale threshold value $T$. Fig. \ref{grayscale threshold} presents the results from our analysis. 
The results 
indicate that the color of weevils significantly differs from other similarly sized objects, making grayscale value differentiation feasible. 
Considering the users' need for sensitivity in weevil detection and allowing a certain margin of error, we set the grayscale threshold $T$ = 60.

\subsection{Experiment 2: Determining the Similarity Threshold S}

To determine the appropriate similarity threshold $S$, we calculated the ratio of the pixel area of the largest possible weevil to the total pixel area of the image. Weevils vary in size, ranging from as small as 1 mm to as large as 38 mm, with most species measuring between 3 to 10 mm. The pea weevil (Bruchus pisorum) typically ranges from 3.9 to 4.9 mm \cite{clement2009pea, animalkingdomWeevilAnimal}. Our system covers a range from 3.5 mm to 18.0 mm, where the maximum weevil size represents an area of 266,000 pixels. 
Given an image resolution of 3856 $\times$ 2490 pixels, we set the similarity threshold $S$ as follows:

\begin{equation*}
S = \left(1 - \frac{266000}{3856 \times 2490}\right) \times 100\% \approx 97.2296\%
\end{equation*}

Thus, we set the similarity threshold $S$ = 97\%. By applying this threshold, we ensure that even the largest weevil entering the device is accurately and individually detected, effectively reducing the impact of dead weevils on detection accuracy and lowering the frequency to clean the device.

In summary, we set the grayscale value threshold $T$ at 60. Contours with areas between 27,785 and 266,000 pixels are considered as weevils: 27,785 $\leq$ Weevil $\leq$ 266,000. If the grayscale value of an independent object is below 60 and its area falls within 27,785---266,000 range, it is identified as a weevil. Mathematically: 
\noindent
\begin{equation*}
\left.
\begin{aligned}
T_{\text{Object}} < 60 \\
27{,}785 \leq \textit{Area}_{\text{Object}} \leq 266{,}000
\end{aligned}
\right\} \quad \text{Object is a Weevil}
\end{equation*}
\noindent

\subsection{Experiment 3: Infrared Sensor Accuracy}
To test the impact of pea weevil size and flying speed on the IR sensor’s accuracy, we conducted experiments using fake weevils of 11 different sizes, mounted on thin threads to control their movement speed. In each set of experiments, we varied only the size or speed, keeping all other variables constant to ensure controlled testing conditions. Consistent lighting was maintained to avoid variations affecting image quality and detection processes. The capture box was placed in a fixed position for all experiments to ensure a consistent background, and images were captured at predetermined intervals for uniform data collection.

\begin{figure}[htbp]
  \centering
  \includegraphics[width=0.45\textwidth]{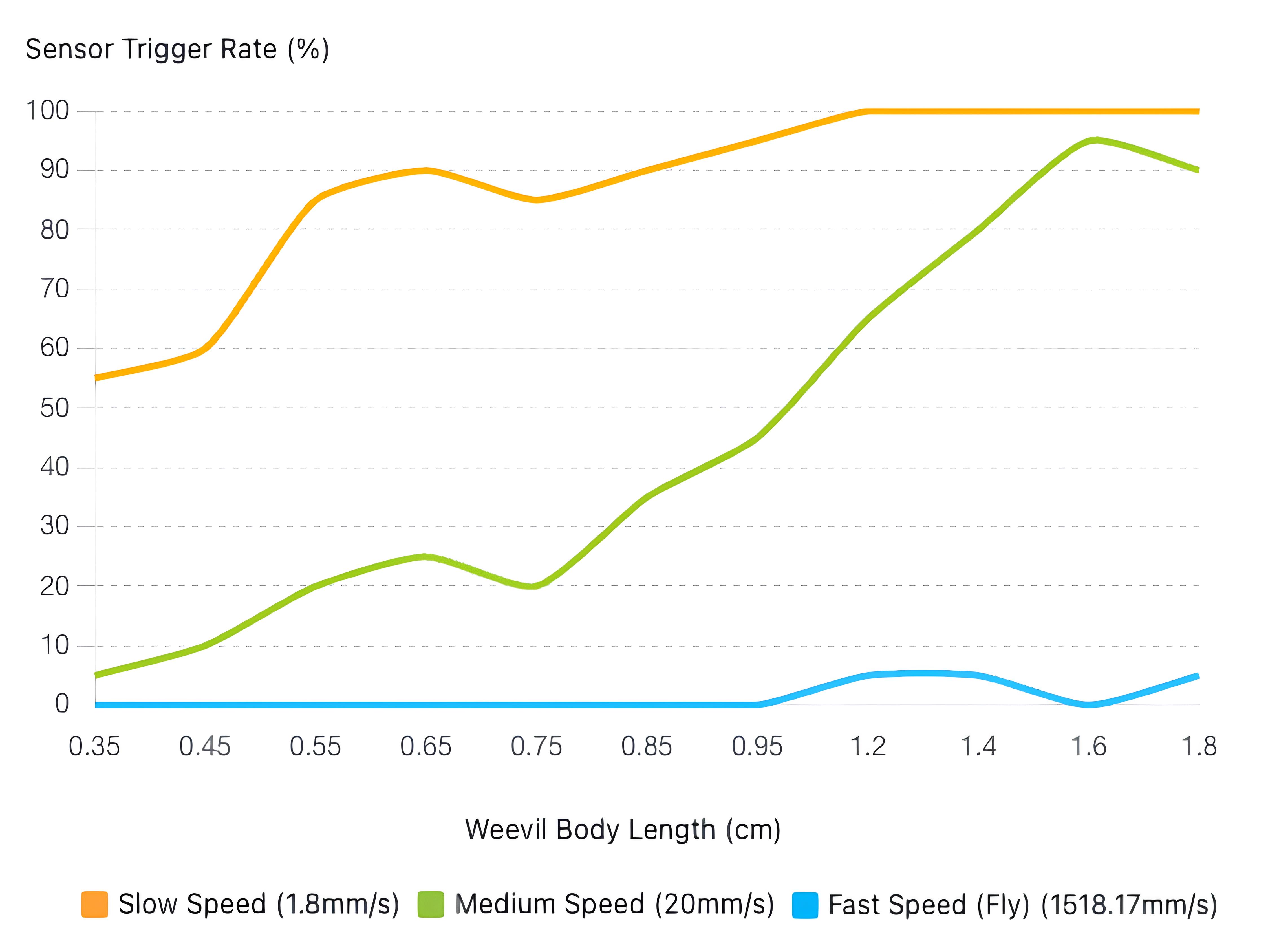}
  \vspace{2pt}
  \caption{\textbf{Slow Speed (1.8 mm/s)}: The sensor shows a high trigger rate, reaching 100\% for weevils larger than 12 mm. \textbf{Medium Speed (20 mm/s)}: The trigger rate increases with weevil size, reaching a maximum of 95\% at 16 mm, indicating that the sensor performs well with larger weevils at this speed. \textbf{Fast Speed (1518.17 mm/s)}: The sensor struggles to detect weevils at high speeds, with trigger rates remaining very low across all sizes, peaking at 5\% for the largest weevils.}
  \label{fig:sensor_trigger_rates}
\end{figure}

We tested the IR sensor’s trigger success rate and the algorithm’s identification success rate using weevils ranging in size from 3.5 mm to 18 mm and movement speeds from 1.8 mm/s to 1518.17 mm/s. The IR sensor, with a refresh rate of once per second, was tested 20 times for each condition, starting with the weevils moving towards the sensor and then away from it. Specifically, we tested the sensor's response as the weevils moved from 110 mm to 40 mm from the sensor, with the sensor set to trigger at distances less than 95 mm.

To simulate realistic weevil movement speeds, we based our speed settings on a study of the red palm weevil's flight performance \cite{mohammed2021design}, which measured a maximum flight speed of approximately 1518 mm/s. We established three speed categories: Slow Speed (1.8 mm/s) to simulate feeding conditions, Medium Speed (20 mm/s) to simulate crawling conditions, and Fast Speed (1518.17 mm/s) to simulate flying conditions. This comprehensive approach allowed us to thoroughly evaluate the system’s performance under various real-world conditions, which is depicted in Fig. \ref{fig:sensor_trigger_rates}.

\subsection{Weevil Detection Results}

\begin{figure*}[htbp]
  \centering
  \includegraphics[width= 16cm]{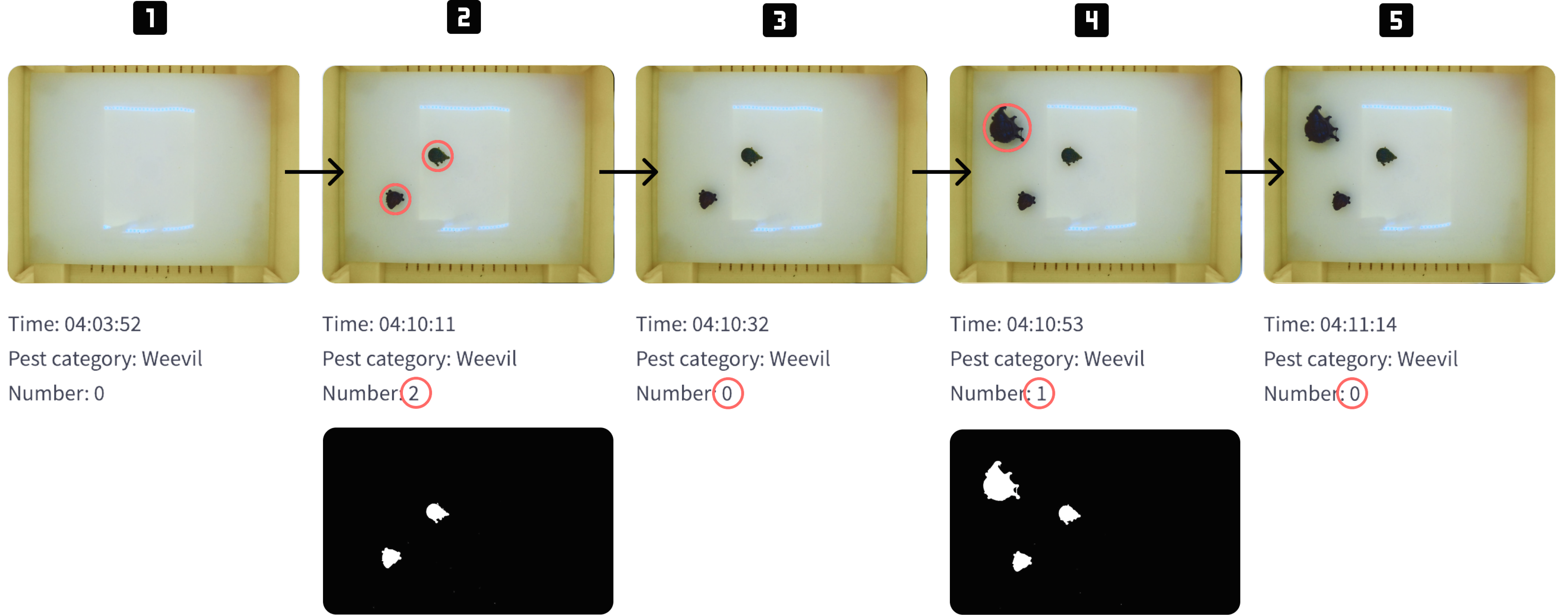}
  \caption{Counting new weevils based on grayscale thresholding}
  \label{fig}
\end{figure*}

Fig. \ref{fig} shows the counting of new weevils based on our threshold-based detection approach, and Table \ref{tab:impact_of_dead_weevils} presents the impact of dead weevils on the detection accuracy.

\subsubsection{Primary Algorithm A}


%

%

Our tests reveal that the system performs well with slower-moving weevils, showing high sensor trigger rates for medium and large sizes. However, the detection rate significantly decreases at higher speeds, indicating a need for optimization in sensor sensitivity or algorithmic adjustments to improve detection under fast-moving conditions. This limitation is from the refresh rate of the IR sensor, which is designed to keep the balance between sensitivity and power consumption.

\subsubsection{Supplemental Algorithm B}

\begin{table*}[htbp]
\caption{Impact of Dead Weevils on Detection Accuracy}
\begin{center}
\begin{tabular}{|c|c|c|c|}
\hline
\textbf{Scenario} & \textbf{Total Tests} & \textbf{Correct Identifications} & \textbf{Accuracy Rate (\%)} \\
\hline
With Dead Weevils & 10 & 10 & 100 \\
\hline
Without Dead Weevils & 10 & 10 & 100 \\
\hline
\end{tabular}
\label{tab:impact_of_dead_weevils}
\end{center}
\end{table*}

With a similarity threshold of 97\%, our tests demonstrated a 100\% accuracy rate in detecting different sizes of dead pea weevils. Additionally, the control experiment, conducted with and without the presence of dead weevils, also achieved 100\% accuracy. These results confirm the algorithm's robustness and effectiveness in accurately distinguishing live weevils, even when dead weevils are present, ensuring reliable performance under various conditions. This finding is crucial for optimizing pest management systems and enhancing user's confidence in the technology.

\section{Discussion}
The developed threshold-based weevil detection system offers several key advantages, including affordability and ease of setup, making it accessible to small-scale farmers. The system's integration with Azure for data storage and processing ensures efficient handling of large data volumes and provides robust remote monitoring capabilities, reducing the need for farmers to be physically present in the fields. However, there are several limitations highlighted by our experiments:
\begin{itemize}
    \item Grayscale Limitations: The grayscale differentiation approach can lead to misidentifications, especially when dealing with objects like dry leaves, soil clumps, and stones. These objects can have grayscale values close to those of weevils, leading to false positives. We only take the average grayscale value from 100 images of each type of interfering object (e.g., soil, leaves). Users can adjust the grayscale threshold based on their specific conditions by capturing their own images.

    \item Algorithm Limitations: We opted not to use machine learning due to the high costs and lack of suitable datasets. Creating a reliable dataset would require extensive time and expense to capture and label numerous weevil images in the field. Additionally, different regions have varying weevil species. While using DSP and computer vision is cost-effective, easy to manage, and quick to set up, it has limitations. For instance, overlapping pests might not be accurately counted, as the system might recognize multiple pests as a single entity. Our algorithm sets the pest body length range between 3.5 mm and 18 mm to include various weevil species. Users can adjust the thresholds based on their crops and pest types, such as moths or other insects.
    
    \item Impact of Dead Weevils: We employ a combination of primary and supplemental algorithms to count weevils. The primary algorithm counts all qualifying objects in the camera's field of view, while the supplemental algorithm focuses on new objects only. Dead insects in the device can lead to false counts if only the primary algorithm is used. The system uses absolute difference to compare consecutive images, applying a 97\% similarity threshold to reduce the impact of dead weevils and lower the need for frequent cleaning.

\end{itemize}
To address these limitations and further validate the system, we plan to conduct field experiments during the pea weevil infestation season in mid-June. Using PLW aggregation pheromones, we aim to test the performance of the system in natural conditions, 
providing insights to refine the detection algorithms and expand the system's applicability to a broader range of pests.

\section{Conclusion}
This paper presents a novel solution in agricultural technology, offering a practical and expandable solution for pest management in resource-constrained environments. By 
promoting sustainable farming practices, the system aims to contribute to the broader mission of alleviating poverty and improving living conditions in disadvantaged communities. Future research will focus on refining the detection algorithms, expanding the range of detectable pests, and developing a machine learning model for categorizing different pest species. Successfully detecting tiny pests like pea weevils will demonstrate our system's accuracy and potentially allow us to expand the system to detect other pests, further enhancing its 
effectiveness in diverse agricultural settings.




\begin{thebibliography}{00}

\bibitem{barbedo2020detecting} J. G. A. Barbedo, ``Detecting and classifying pests in crops using proximal images and machine learning: A review,” Ai, vol. 1, no. 2, pp. 312–328, 2020.

\bibitem{reddy2018pheromone} G. V. Reddy, G. Shrestha, D. A. Miller, and A. C. Oehlschlager, ``Pheromone-trap monitoring system for pea leaf weevil, sitona lineatus: Effects of trap type, lure type and trap placement within fields,” Insects, vol. 9, no. 3, p. 75, 2018.

\bibitem{mohammed2021design} M. Mohammed, H. El-Shafie, and N. Alqahtani, ``Design and validation of computerized flight-testing systems with controlled atmosphere for studying flight behavior of red palm weevil, rhyn-chophorus ferrugineus (olivier),” Sensors, vol. 21, no. 6, p. 2112, 2021.

\bibitem{st2018evaluation} A. St Onge, H. A. C´arcamo, and M. L. Evenden, ``Evaluation of semiochemical-baited traps for monitoring the pea leaf weevil, sitona lineatus (coleoptera: Curculionidae) in field pea crops,” Environmental entomology, vol. 47, no. 1, pp. 93–106, 2018.

\bibitem{farooq2021iot} M. S. Farooq and S. Akram, ``Iot in agriculture: challenges and opportunities.” Journal of Agricultural Research (03681157), vol. 59, no. 1, 2021.

\bibitem{reddy2010new} G. V. Reddy and A. Guerrero, ``New pheromones and insect control strategies,” Vitamins \& hormones, vol. 83, pp. 493–519, 2010.

\bibitem{patricio2018computer} D. I. Patr´ıcio and R. Rieder, ``Computer vision and artificial intelligence in precision agriculture for grain crops: A systematic review,” Computers and electronics in agriculture, vol. 153, pp. 69–81, 2018.

\bibitem{tian2020computer} H. Tian, T. Wang, Y. Liu, X. Qiao, and Y. Li, ``Computer vision technology in agricultural automation—a review,” Information Processing in Agriculture, vol. 7, no. 1, pp. 1–19, 2020.

\bibitem{jung2021deep} D.-H. Jung, N. Y. Kim, S. H. Moon, C. Jhin, H.-J. Kim, J.-S. Yang, H. S. Kim, T. S. Lee, J. Y. Lee, and S. H. Park, ``Deep learning-based cattle vocal classification model and real-time livestock monitoring system with noise filtering,” Animals, vol. 11, no. 2, p. 357, 2021.

\bibitem{githubGitHubKrantLeeee515MicrosoftFarmBeats} ``515-Microsoft-FarmVibes--- TECHIN-515 Project GitHub Repository,” \href{https://github.com/KrantLeeee/515-Microsoft-FarmVibes-FarmSentinel}{github.com/KrantLeeee/515-Microsoft-FarmVibes-FarmSentinel}, [Accessed 08-08-2024].

\bibitem{abrar2023anomaly} M. M. Abrar and S. Hariri, ``An anomaly behavior analysis framework for securing autonomous vehicle perception,” in 2023
20th ACS/IEEE International Conference on Computer Systems and Applications (AICCSA). IEEE, 2023, pp. 1–6.

\bibitem{animalkingdomWeevilAnimal} ``Weevil—Animal Kingdom,” \href{https://animalkingdom.org/species/weevil/}{animalkingdom.org/species/weevil/},
[Accessed 10-08-2024].

\bibitem{durgabai2018pest} R. Durgabai, P. Bhargavi et al., ``Pest management using machine learning algorithms: a review,” International Journal of Computer
Science Engineering and Information Technology Research (IJCSEITR), vol. 8, no. 1, pp. 13–22, 2018

\bibitem{dore1995farm} T. Dor´e and J.-M. Meynard, ``On-farm analysis of attacks by the pea weevil (sitona lineatus l.; col., curculionidae) and the resulting damage to pea (pisum sativum l.) crops,” Journal of Applied Entomology, vol. 119, no. 1-5, pp. 49–54, 1995.

\bibitem{clement2009pea} S. Clement, K. McPhee, L. Elberson, and M. Evans, “Pea weevil, bruchus pisorum l.(coleoptera: Bruchidae), resistance in pisum
sativum $\times$ pisum fulvum interspecific crosses,” Plant Breeding, vol. 128, no. 5, pp. 478–485, 2009.

\bibitem{meisner2016data} M. H. Meisner, J. A. Rosenheim, and I. Tagkopoulos, ``A data-driven, machine learning framework for optimal pest management in cotton,” Ecosphere, vol. 7, no. 3, p. e01263, 2016.

\bibitem{huang2022deep} X. Huang, Z. Hu, Y. Qiao, and S. Sukkarieh, ``Deep learning-based cow tail detection and tracking for precision livestock farming,” IEEE/ASME Transactions on Mechatronics, 2022.

\end{thebibliography}


\end{document}